\documentstyle[12pt,epsfig]{article} 
\begin{document}
\vspace{-1cm} 
\begin{flushright}
\begin{tabular}{l}
   OCHA-PP-109
\end{tabular}
\end{flushright}
\vspace{2mm}
\baselineskip24pt
\begin{center}
{\Large \bf Determination of parameters of Higgs sector in the

minimal supersymmetric Standard Model 
}\\
\vspace{5mm}
\baselineskip18pt
{\bf Jun-ichi Kamoshita}\\

{\it Department of Physics of Ochanomizu University,\\
     2-1-1, Otsuka, Bunkyo, Tokyo 112, Japan}\\
\vspace{10mm}
{\bf ABSTRACT} \\
\baselineskip18pt
\vspace{5mm}
\begin{minipage}{14.5cm}
      We examine whether parameters related to Higgs sector of the 
      minimal supersymmetric standard model can be determined by
      detailed study of production cross section and decay branching
      ratios of the Higgs boson. 
      Assuming that only the light Higgs boson is observed at a 
      future $e^+e^-$ linear collider with $\sqrt{s}=300\sim500$GeV,
      we show that values of $m_{susy}$ and $\tan\beta$ are restricted within
      a narrow limits in the $m_{susy}$ versus $\tan\beta$ plane by the
      combined analysis of the light Higgs properties.
      It is also pointed out that, in some case, 
      the value of $\tan\beta$ may be restricted
      within a relatively low value, $\tan\beta=1\sim5$.
\end{minipage}

\end{center}
\vspace{10mm} 
\baselineskip16pt

 In the search of the theory beyond the standard model,
 supersymmetric(SUSY) extension of the minimal standard model(SM) is
 considered to be an attractive and promising candidate.
 It is, therefore, important to investigate 
 how the idea of SUSY can be explored in 
 the future collider experiments
 such as LHC and $e^+e^-$ linear colliders.
 In this respect, the Higgs sector of the SUSY standard models can 
 play a unique role. Since the Higgs sector has distinct features,
 its close investigation can give important information on 
 the structure of these models.

 In the minimal supersymmetric standard model(MSSM) the Higgs sector
 consists of two Higgs doublets, therefore, 
 there exist five physical states; 
 i.e. two CP-even neutral Higgs bosons($h$ and $H$ with $m_h<m_H$),
 one CP-odd Higgs boson$(A)$, and 
 one pair of charged Higgs bosons$(H^\pm)$.
 It is possible to derive specific 
 predictions for this Higgs sector because
 the form of Higgs potential in the MSSM is very restricted
 in comparison with that in the general two Higgs doublets model.
 Especially, the upper bound on the mass of lightest CP-even neutral
 Higgs boson is given as about 130GeV\cite{OYY}.
 As for the possibility of Higgs boson discovery,
 it have been shown that at least one of
 CP-even neutral Higgs bosons is detectable at future $e^+e^-$
 linear collider with $\sqrt{s}=300\sim500$GeV\cite{Janot}.
 Furthermore, the detectability of Higgs boson 
 is guaranteed for large class of 
 SUSY models with extended Higgs sector\cite{Kamo}.

 After the discovery of the Higgs boson in the MSSM,
 one of the questions of interest is what extent the parameters
 relating to the Higgs sector will be constrained from 
 the detailed study of properties of the Higgs boson.
 By precisely measured branching ratios of the Higgs boson,
 the mass of CP-odd Higgs boson($m_A$) can be constrained 
 with almost independent of SUSY breaking mass scale($m_{susy}$)
 even when the CP-odd Higgs boson will not be discovered 
 at future linear colliders with $\sqrt{s}=300$GeV\cite{KOT}.
 In this paper we consider the determination of
 parameters of Higgs sector in the MSSM assuming that 
 only the lightest CP-even neutral Higgs boson will be observed
 at the future $e^+e^-$ linear collider with $\sqrt{s}=300\sim500$GeV.
 It is shown that the $m_{susy}$-$\tan\beta$ parameter space
 can be restricted within narrow region by 
 precise measurements of Higgs boson properties.

 Let us begin by listing the parameters of Higgs sector and 
 the observables which can be used to determine these parameters.
 At the tree level, the masses of Higgs bosons and the mixing 
 among Higgs bosons are parametrized by two parameters, 
 i.e. CP-odd Higgs boson mass and the ratio of 
 the vacuum expectation values 
 ($\tan\beta=\frac{\langle H_2 \rangle}{\langle H_1 \rangle}$), where 
 the $H_1$ is a Higgs doublet that couples to up-type quarks and 
 the $H_2$ is a Higgs doublet that couples to down-type quarks 
                                           and leptons.
 However, once the radiative corrections to the 
 Higgs potential is taken into account, 
 they bring new parameters in our analysis.
 In the calculation of the Higgs effective potential 
 at one loop level,
 most important contribution comes from top and stop loop and
 therefore the relevant parameters are 
 two stop masses$(m_{\tilde{t_1}}, m_{\tilde{t_2}})$,
 Higgsino mass parameter$(\mu)$ and 
 trilinear soft-breaking parameter$(A_t)$.
 For the moment, we assume that no significant effect is 
 induced from the left-right mixing of the stop sector.
 Then, effectively 
 there are three parameters relating to the Higgs sector.
 Usually,  for these three parameters, 
 we take $m_A$, $\tan\beta$, and  $m_{susy}$ defined by
 $m_{susy}=\sqrt{m_{\tilde{t_1}}m_{\tilde{t_2}}}$.
 Then the CP-even Higgs mass matrix is 
\begin{eqnarray}
   M^2_{higgs}&=&
       \left(\begin{array}{cc} 
         m^2_Z\cos^2\beta+m^2_A\sin^2\beta 
                             & -(m^2_Z+m^2_A)\cos\beta\sin\beta\cr
        -(m^2_Z+m^2_A)\cos\beta\sin\beta &
                  m^2_Z\sin^2\beta+ m^2_A\cos^2\beta 
                                + \frac{\delta_t}{\sin^2\beta}
                        \end{array}\right) ,
\end{eqnarray}
where 
\begin{eqnarray}
   \delta_t=\frac{3m_t^4}{4\pi^2v^2}
                  \ln\left(\frac{m^2_{susy}}{m^2_t}\right) 
\end{eqnarray}
 represents the leading part of the radiative corrections 
 stem from the top-stop loop effect. 
 The mass of neutral Higgses and
 Higgs mixing angle, $\alpha$, are 
 given by
\begin{eqnarray} 
   m^2_h&=& \frac{1}{2}\left[ 
                  m^2_A+m^2_Z+\delta_t/\sin^2\beta
                      \right. \nonumber \\
        & & \left. 
         -\sqrt{\{(m^2_Z-m^2_A)\cos2\beta-\delta_t/\sin^2\beta\}^2
                    +(m^2_Z+m^2_A)^2\sin^22\beta }
            \right]\label{eqn:mh} ,\\
    m^2_H&=& m^2_A+m^2_Z-m^2_h+\delta_t/\sin^2\beta ,\\
 \tan\alpha&=&\frac{(m^2_Z+m^2_A)\cos\beta\sin\beta}
                 {m^2_h-(m^2_Z\cos^2\beta+m^2_A\sin^2\beta)} \ .
\end{eqnarray}

 The lightest CP-even neutral Higgs boson is mainly produced through
 the Higgs bremsstrahlung process, $e^+e^-\to Zh$, 
 at $e^+e^-$ linear collider with $\sqrt{s}=300\sim500$GeV.
 If we assume that the decay modes of the Higgs boson to SUSY particles
 are not dominant\footnote{ 
             If the decay mode of Higgs boson to 
          the SUSY particles will be observed, 
          we can see obviously that 
          the Higgs boson belongs to SUSY model.
          We are now interesting in the case that 
          the SM-like Higgs boson will be observed.
          Therefore we will not consider about such a case.
          },
 then the main decay mode of the Higgs boson is 
 the $h\to b\bar{b}$ mode.
 In this case, 
 the behavior of the Higgs boson may be similar to 
 that of Higgs boson in the SM.
 The lightest Higgs boson then has sizable decay branching ratios
 in the modes $h\to b\bar{b}, \tau\bar{\tau}, c\bar{c}$
 and $gg$.\footnote{Since availability of $h\to WW^{\ast}$ depends 
                    crucially on the Higgs boson mass, we will not 
                    consider this mode here.}

 With a reasonable luminosity of $\sim50$fb$^{-1}$/year, 
 the mass of the Higgs boson, $m_h$, can be determined precisely
 by the recoil mass distribution\cite{JLC,WCHB,NLCHWG}.
 The Higgs production cross section,
 $\sigma(e^+e^-\to Zh)$, is obtained by
 the branching ratio of $Z$ boson decaying to $l\bar{l}(l=e, \mu)$ and 
 the cross section of the event with 
 the recoil mass around $m_h$\cite{WCHB}.
 The production cross section multiplied by 
 the branching ratio of $h\to X (X=\{b\bar{b}\}, \{\tau\bar{\tau}\}, 
 \{c\bar{c}$~{\scriptsize or}~$gg\})$\footnote{Although 
               it is very difficult to measure the branching ratios of
               the mode $c\bar{c}$ and $gg$ separately,
               the sum of $Br(h\to c\bar{c})$ and $Br(h\to gg)$ can be
               measured with reasonable
               precision\cite{WCHB,NLCHWG,HBB,Nakamura}.
               We denote the sum of $Br(h\to c\bar{c})$ and $Br(h\to gg)$ as 
               $Br(h\to c\bar{c}\ { }_{\rm or}\ gg)$.},
 $\sigma(e^+e^-\to Zh)Br(h\to X)$,
 can be obtained by the $ZX$ production rate with 
 the invariant mass of $X$ to be around $m_h$\cite{WCHB,NLCHWG}.

 The ratio of branching ratios or the ratio of partial decay widths
 are obtained by 
\begin{eqnarray}
    \frac{Br(h\to X_1)}{Br(h\to X_2)}
   =\frac{\Gamma(h\to X_1)}{\Gamma(h\to X_2)}
   =\frac{\sigma(e^+e^-\to Zh)Br(h\to X_1)}
         {\sigma(e^+e^-\to Zh)Br(h\to X_2)},
\end{eqnarray}
 where $X_1$ and $X_2$ are $\{b\bar{b}\}, \{\tau\bar{\tau}\}$ or 
 $\{c\bar{c}$~{\scriptsize or}~$gg\}$.
 We can expect that 
 $Br(h\to~\tau\bar{\tau})/Br(h\to~b\bar{b})$ and 
 $Br(h\to~c\bar{c}$~{\scriptsize or}~$gg)/Br(h\to~b\bar{b})$
 will be determined in a reasonable precision\cite{WCHB,NLCHWG,Nakamura}.

 The formulas for partial decay width of 
 Higgs boson in MSSM is found, for example, in \cite{Barger}. 
 Higgs-fermion-fermion couplings are listed in 
 Table~\ref{tb:cpl}.
 The partial decay width for 
 $h\to b\bar{b}$ and $h\to\tau\bar{\tau}$
 are proportional to the down-type fermion-Higgs coupling, and then
 the ratio $Br(h\to~\tau\bar{\tau})/Br(h\to~b\bar{b})$ is 
 the same as that in the SM. Therefore no information on 
 the parameters of Higgs sector in the MSSM are obtained from 
 this ratio.\footnote{
                    This ratio is important to determine the bottom mass
                    as discussed, for example, in\cite{KOT,NLCHWG}.}
 On the other hand, as reported in \cite{KOT}, 
 the ratio of 
 $Br(h\to c\bar{c}$~{\scriptsize or}~$gg)/Br(h\to b\bar{b})$
 is useful variable to constrain the value of $m_A$,
 because $Br(h\to c\bar{c}$~{\scriptsize or}~$gg)/Br(h\to b\bar{b})$ 
 depend on $m_A$ strongly and almost independent of $m_{susy}$.

\begin{table}[t]
\begin{center}
\begin{tabular}{ c|c|c|c }
        & h-u-u & h-d-d & h-l-l \\
\hline
        & & & \\

 MSSM   & $-i\frac{\mbox{$m_u$}}{\mbox{$v$}}\frac{\cos\alpha}{\sin\beta}$ 
        & $ i\frac{\mbox{$m_d$}}{\mbox{$v$}}\frac{\sin\alpha}{\cos\beta}$ 
        & $ i\frac{\mbox{$m_l$}}{\mbox{$v$}}\frac{\sin\alpha}{\cos\beta}$ \\
%\hline
        & & & \\
  SM    & $-i\frac{\mbox{$m_u$}}{\mbox{$v$}}$ 
        & $-i\frac{\mbox{$m_d$}}{\mbox{$v$}}$ 
        & $-i\frac{\mbox{$m_l$}}{\mbox{$v$}}$ \\
\end{tabular}
\end{center}
\begin{center}
\begin{minipage}{13cm}\caption[]{\small
         Couplings of light Higgs boson to fermion pair in 
         the MSSM and the SM.
         The $u,d$ and $l$ stand for up-type quarks,
         $u=\{u,c,t\}$; down-type quarks, $d=\{d,s,b\}$;
         leptons, $l=\{e,\mu,\tau\}.$ 
            }\label{tb:cpl}
\end{minipage}
\end{center}
\end{table}

\begin{figure*}\vspace{-0.3cm}
\epsfxsize = 14 cm
\centerline{\epsfbox{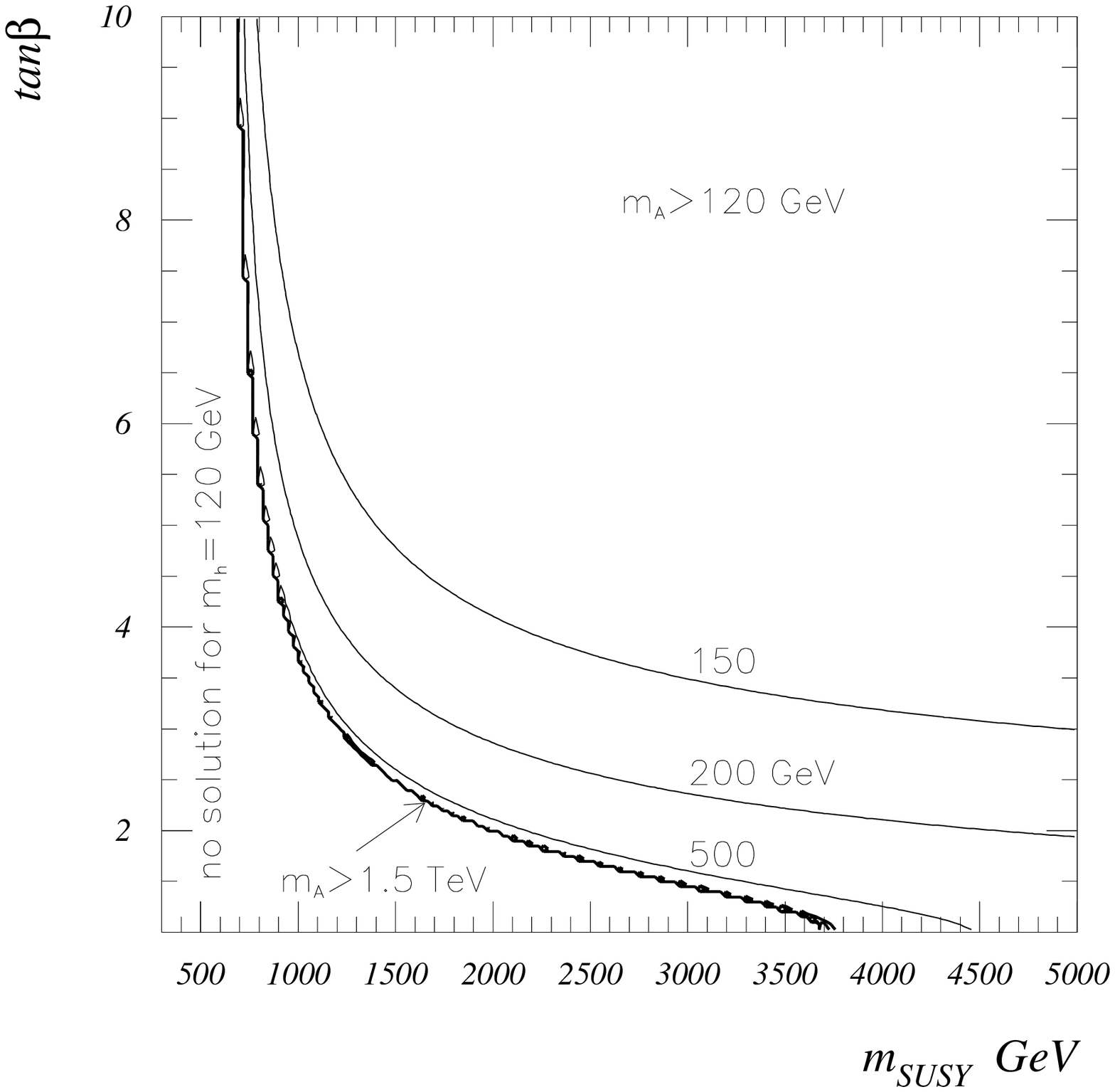}}
\caption[]{   }
\label{fign:ma}
\end{figure*}

 The three parameters; 
 $m_A,\tan\beta$ and $m_{susy}$;
 will be restricted by the observables mentioned above.
 The errors of observables
 have been estimated in detail\cite{NLCHWG}.
 According to their estimation, 
 the error of $m_h$ is to be 0.1$\sim$0.5\%. 
 Therefore, in the following,
 we will treat the $m_h$ as fixed variable.
 Then one of the three parameters is derived from 
 other two with the formula of the lightest Higgs boson
 mass(\ref{eqn:mh}).
 Thus the remaining degrees of freedom of parameters are two.
 Hereafter we choose $m_{susy}$ and $\tan\beta$ as free parameters 
 and then derive the value of $m_A$ with Higgs mass formula 
 for the fixed value of $m_h$.
 For $m_h=120$GeV, figure~\ref{fign:ma} shows the contour plot of $m_A$ 
 in the $m_{susy}$ versus $\tan\beta$ plane.

 We can constrain the value of $m_A$ by
 $Br(h\to c\bar{c}$~{\scriptsize or}~$gg)/Br(h\to b\bar{b})$
 \cite{KOT}. 
 The knowledge of the magnitude of $m_A$ is useful to set
 the center of mass energy at the next step of 
 future $e^+e^-$ linear collider experiments. 

 However, determination of $m_{susy}$ is 
 necessary to know SUSY breaking scale.
 Determination of $\tan\beta$ 
 has great impact on both theoretical and 
 experimental study of SUSY model,
 because not only the physics of Higgs sector 
 but also that of other
 SUSY sector, for example chargino and neutralino sector, 
 depend on $\tan\beta$. 
 Therefore we must start to use other observables
 in order to determine the value of both $m_{susy}$ and $\tan\beta$.
 
\begin{figure*}\vspace{-0.3cm}
\epsfxsize = 14 cm
\centerline{\epsfbox{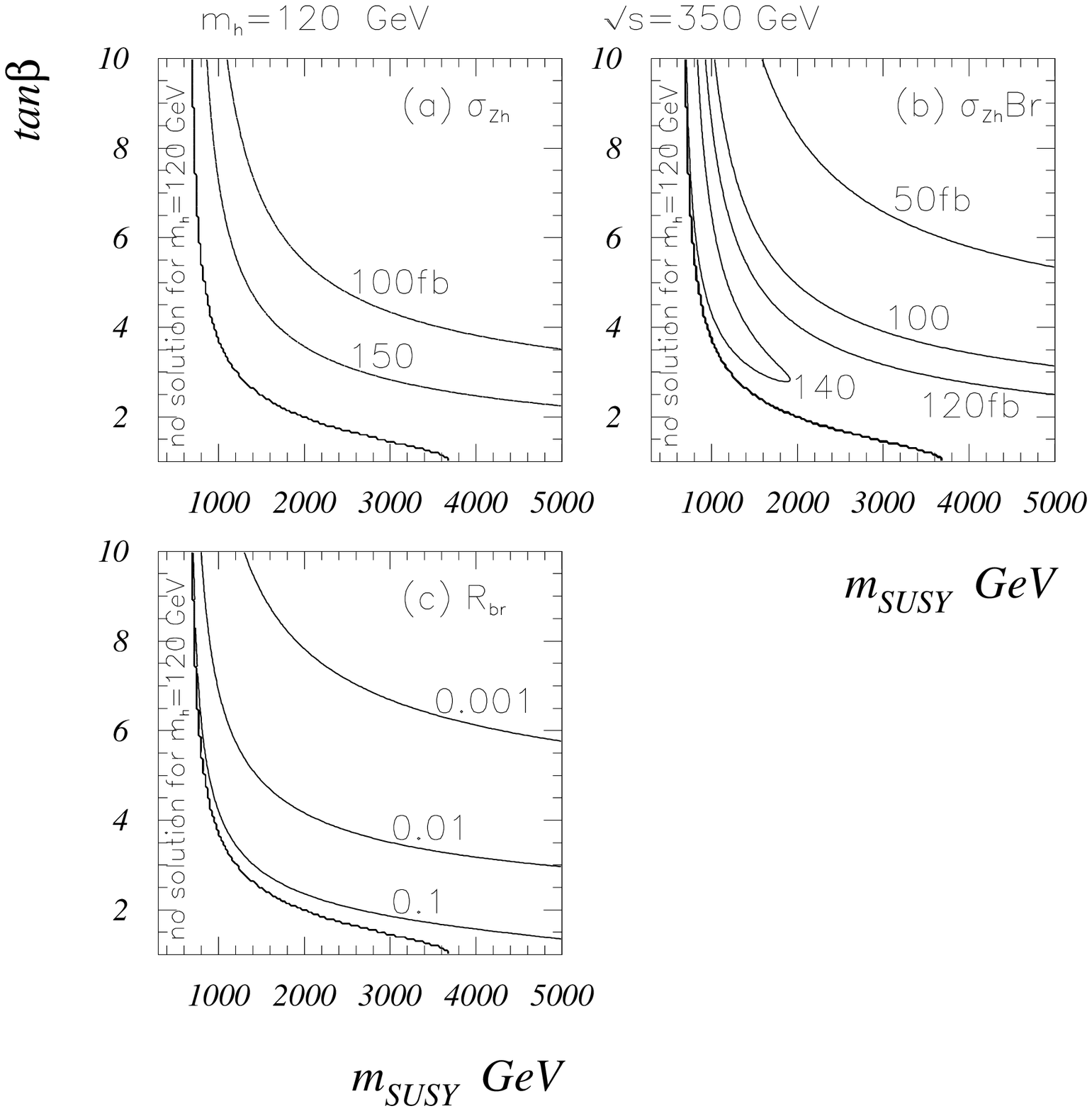}}
\caption[]{   }
\label{fign:obs}
\end{figure*}

 Both $\sigma(e^+e^-\to Zh)$ and 
 $\sigma(e^+e^-\to Zh)Br(h\to~b\bar{b})$ depend on 
 the angle $\alpha$ and $\beta$ as follows,
 \begin{eqnarray}
  \sigma(e^+e^-\to Zh)&\propto&\sin^2(\alpha-\beta) , \nonumber\\
  \sigma(e^+e^-\to Zh)Br(h\to b\bar{b})&\propto&
    \sin^2(\alpha-\beta)\left(\frac{\sin\alpha}{\cos\beta}\right)^2
    \nonumber .
 \end{eqnarray}
 As discussed in \cite{KOT}, 
 we obtain the following approximate relation,
 \begin{eqnarray}
  \frac{Br(h\to c\bar{c}\ { }_{\rm or}\ gg)}{Br(h\to b\bar{b})}
                 \propto\left(\frac{1}{\tan\beta\tan\alpha}\right)^2
  \nonumber. 
 \end{eqnarray}
 Therefore these observables give us 
 different constraint on the value of
 $m_{susy}$ and $\tan\beta$.
 Hereafter we use shortened notations defined as follows,
 $\sigma_{Zh}\equiv\sigma(e^+e^-\to Zh)$,\\
 $\sigma_{Zh}Br(b\bar{b})\equiv\sigma(e^+e^-\to Zh)Br(h\to b\bar{b})$
 and 
  $R_{br}\equiv
  Br(h\to c\bar{c}$~{\scriptsize or}~$gg)/Br(h\to b\bar{b})$.
 Figure~\ref{fign:obs}(a)(b)(c) show the contour plots of 
 $\sigma_{Zh}$, $\sigma_{Zh}Br(b\bar{b})$ and $R_{br}$ respectively
 in the $m_{susy}$ versus $\tan\beta$ plane
 when $m_h=120$GeV.
 The shape of contours in figure \ref{fign:obs}(b)
 is somewhat different from the other two
 in the left side of the figure.
 Figure~\ref{fign:obs}(a) of $\sigma_{Zh}$ is similar to
 figure~\ref{fign:obs}(c) of $R_{br}$,
 however, figure~\ref{fign:obs}(a) shows gentle slope
 as compared to figure~\ref{fign:obs}(c).

 Now we combine these observables to estimate for the constraints on 
 the values of $m_{susy}$ and $\tan\beta$.
 For this purpose, 
 we take $m_{susy}$ and $\tan\beta$ as fitting parameters 
 and then  perform the $\chi^2$ test in 
 the $m_{susy}$ versus $\tan\beta$ plane 
 for the fixed value of $m_h$, for example $m_h=120$GeV.
 As for the value of $m_A$, its value is derived from 
 Higgs mass formula(\ref{eqn:mh})
 at point by point in the $m_{susy}$ versus $\tan\beta$ plane.

 However, we input the values of $m_h$, $m^0_A$ and $m^0_{susy}$ 
 as true values for the $\chi^2$
 test\footnote{
       In order to distinguish "true" value of  $m_A$ and $m_{susy}$
       from the mass of $m_A$ and $m_{susy}$ as fitting parameters, 
       the index "$^0$" is appended to the "true" variables.
                },
 because these variables have clear physical meanings as 
 a mass of particles or a typical mass scale for $m_{susy}$.
 As for $\tan\beta$, the "true" value is calculated from
 the input parameters; $m_h$, $m^0_A$ and $m^0_{susy}$; 
 by Higgs mass formula(\ref{eqn:mh}).
 
%
% Estimated Errors of Observable discussed in NLCHWG.
% 
\begin{table}[t]
\begin{center}
\begin{tabular}{|c|c|c|c|c|} 
\hline
  $m_h$   & $\delta(m_h)$ & $\delta(\sigma_{Zh})$ 
                  & $\delta(\sigma_{Zh}Br(b\bar{b}))$ 
                  & $\delta(R_{br})$  \\
\hline
  110 GeV & $0.1\sim 0.5$\% & $\sim 7$\% & $\sim 2.5$\% &$\sim 14$\% \\
  120 GeV & $0.1\sim 0.5$\% & $\sim 7$\% & $\sim 3.5$\% &$\sim 14$\% \\
\hline
\end{tabular}
\end{center}
\begin{center}
\begin{minipage}{13cm}\caption[]{\small
     List of error for observables discussed in \cite{NLCHWG}. 
     $R_{br}$ is defined by
$R_{br}\equiv Br(h\to c\bar{c}$~{\scriptsize or}~$gg)/Br(h\to b\bar{b})$
\cite{KOT}.
            }\label{tb:error}
\end{minipage}
\end{center}
\end{table}

 Definition of $\chi^2$ is given by
\begin{eqnarray}
 \chi^2\equiv&&\!\!\!\!\!\!\!
  \left\{\left(\frac{\sigma_{Zh}-\sigma_{Zh}^{0}}
                               {\delta\sigma_{Zh}}\right)^2 
     +\left(\frac{\sigma_{Zh}Br(b\bar{b})
                     -\sigma_{Zh}Br(b\bar{b})^{0}}
                         {\delta(\sigma_{Zh}Br(b\bar{b}))}\right)^2 
     + \left(\frac{R_{br} - R_{br}^{0}}{\delta(R_{br})}\right)^2
             \right\} .
  \label{eqn:chisq}
 \end{eqnarray} 
 The $\delta(\sigma_{Zh})$, $\delta(\sigma_{Zh}Br(b\bar{b}))$ and 
 $\delta(R_{br})$ are the expected experimental errors.
 The estimated error of each observable reported in \cite{NLCHWG}
 is summarized in Table \ref{tb:error}.
 The $\sigma_{Zh}^{0}$, $\sigma_{Zh}Br(b\bar{b})^{0}$ and 
 $R_{br}^{0}$ are the central value 
 derived from the input parameters, $m_h, m_A^0$ and $m_{susy}^0$.
 The $\sigma_{Zh}$, $\sigma_{Zh}Br(b\bar{b})$ and $R_{br}$ are 
 the variables calculated at each point in the  
 $m_{susy}$ versus $\tan\beta$ plane.
 To calculate Higgs production cross section,
 we use $\sqrt{s}=350$GeV.\footnote{
              Of course when $m_A<\sqrt{s}/2$,
        the CP-odd Higgs boson will be produced by 
        associated production process, $e^+e^-\to AH$.
        In this case we can use many observables depending on 
        SUSY parameters and 
        should convert a strategy of our analysis to another one. 
        Since we assume that only light Higgs boson is discovered,
        we constrain our analysis for $m_A>\sqrt{s}/2\sim 180$GeV.
        Hereafter we will not consider the case for $m_A<180$GeV. 
              }

 The contour plots of $\chi^2$ for $m_h=120$GeV are shown in 
 Figure~\ref{fign:ch2i}(a)(b) with 95\%CL contour. 
 We find in figure~\ref{fign:ch2i}(a) that the $\tan\beta$ is restricted 
 within a relatively small value, $\tan\beta<4.5$, and 
 the value of $m_{susy}$ is weakly restricted, $m_{susy}>1$TeV.
 Figure~\ref{fign:ch2i}(b) shows the contour plot of $\chi^2$ 
 for other input value.
 In figure~\ref{fign:ch2i}(b), 
 although  the upper bound on the $m_{susy}$ and $\tan\beta$ 
 is not restricted in the displayed region,
 the $m_{susy}$-$\tan\beta$ parameter space is restricted within
 a narrow limits.

 This results can be understood as follows.
 Once the value of $m_h$ is fixed, $\tan\beta$ and $m_{susy}$ is 
 strongly correlated by Higgs mass formula(3). 
 We can consider that figure~\ref{fign:ma} shows 
 the value of $\tan\beta$ as a function of $m_{susy}$
 for fixed values of $m_h$ and $m_A$. From figure~\ref{fign:ma},
 the $m_{susy}$-$\tan\beta$ parameter space is restricted within
 a relatively narrow region 
 even if $m_A$ vary from  $\sim 200$GeV to larger than 1 TeV.
 However the constraint obtained from figure~\ref{fign:ma}
 is somewhat weak as compared to figure~\ref{fign:ch2i}(a)(b).
 Figure~\ref{fign:ch2spp} show the contour of $\chi^2$ superposed 
 by the curves for central value of each observable.
 We can see from figure~\ref{fign:ch2spp} that the $R_{br}$
 contribute strongly to the constraint on 
 the $m_{susy}$-$\tan\beta$ plane.
 In figures~\ref{fign:ch2i}(a)~and~\ref{fign:ch2spp},
 the value of $m_A$ is restricted within 
 about 180$\sim$230GeV by $R_{br}$ and as a result
 the region satisfying the constraints become narrow
 as compared to that obtained by figure~\ref{fign:ma}.
 The reason why the upper bound on $\tan\beta$ is obtained in 
 figure~\ref{fign:ch2i}(a) is, in addition to $R_{br}$, 
 the $\sigma_{Zh}Br(b\bar{b})$ contribute effectively to 
 the constraint on the $m_{susy}$-$\tan\beta$ plane. 

      So far, we have neglected the L-R mixing of stop sector. 
      In case of taking account of the L-R mixing effect,
      contours in figures should be shifted with 
      varying the value of $A_t$ and $\mu$ parameter.
      However 
      the result in our analysis will not 
      change essentially, because
      $R_{br}$ is almost independent of 
      the parameters of stop sector as shown in \cite{KOT}. 

%==================================================== 
 On the theoretical aspect, 
 the requirement of yukawa coupling unification in 
 SUSY-GUT\cite{YKWU} restricts the value of $\tan\beta$ within 
 two solutions.
 One is the small $\tan\beta$ solution, $\tan\beta=1\sim 3$. 
 Another one is the large $\tan\beta$ solution, $\tan\beta\sim 50$.
 There are mainly two types of scenarios for 
 yukawa coupling unification.
 These are the bottom-tau yukawa unification and 
 the top-bottom-tau yukawa unification scenarios.
 The requirement of bottom-tau yukawa coupling unification suggests
 both the small $\tan\beta$ solution and the large $\tan\beta$ solution. 
 However the requirement of top-bottom-tau yukawa coupling 
 unification suggests only the large $\tan\beta$ solution.
 Therefore, if large value of $\tan\beta$ will be excluded by 
 precise measurements of light Higgs properties at the 
 future linear collider,
 as we have shown in case of figure~\ref{fign:ch2i}(a),
 the experiments may rule out
 the top-bottom-tau yukawa unification scenario
 even when only the lightest Higgs boson will be observed.
 
 We conclude our discussion.
 We have examined 
 whether the parameters of Higgs sector in the MSSM
 can be determined by detailed study of Higgs properties.
 We show that 
 the value of $\tan\beta$ and $m_{susy}$ is restricted
 within the very narrow limits even when 
 only the light Higgs boson is discovered 
 and its mass is determined precisely.
 The $R_{br}$ contributes largely to the constraint on
 the $m_{susy}$-$\tan\beta$ plane.
 We also show that, in some case, 
 the upper bound on $\tan\beta$ may be obtained  
 by combining analysis of observables such as 
 $\sigma_{Zh}$, $\sigma_{Zh}Br(b\bar{b})$ and $R_{br}$.
 However to obtain more strict constraint on 
 both $m_{susy}$ and $\tan\beta$, 
 we have need of constraints by other quantities
 obtained from heavy Higgs and/or SUSY particles.  

 The author would like to thank A. Sugamoto and M. Aoki for
 reading the manuscript and helpful comments.

\newpage 
\begin{center}
    {\large Figure Caption}
\end{center}
\begin{description}
\item[Figure 1:] Contour plot of the $m_A$ for $m_h=120$GeV are 
        shown in the $m_{susy}$ versus $\tan\beta$ plane.
        We take the top quark mass as $m_t=175$GeV.
        The Higgs mass formula(\ref{eqn:mh}) can not be satisfied 
        for $m_h=120$GeV in
        the left and bottom left region of figure. 
        In the large $\tan\beta$ and large $m_{susy}$ region,
        the value of $m_A$ is always larger than 120GeV 
        which is the value of $m_h$.
\item[Figure 2:] Contours plot of (a) $\sigma_{Zh}$; 
        (b) $\sigma_{Zh}Br(b\bar{b})$; and (c) $R_{br}$ are shown.
        We take the quark masses as 
        $m_t=175$GeV, $m_b(m_b)=4.2$GeV and $m_c(m_c)=1.3$GeV.
        The strong coupling constant is taken as
        $\alpha_s(m_Z)=0.12$.
\item[Figure 3:]  Contour plot of $\chi^2$ with $\chi^2=4.61$
        (a) for 
        ($m_h, m^0_A, m^0_{susy}$)=(120GeV, 200GeV, 3500GeV) and 
        (b) for 
        ($m_h, m^0_A, m^0_{susy}$)=(120GeV, 250GeV, 1000GeV).
         $\chi^2<4.61$ for the inside of a narrow region.
\item[Figure 4:]  Contour plot of $\chi^2$ with $\chi^2=4.61$;
        curves giving the central values of $\sigma_{Zh}$, 
        $\sigma_{Zh}Br(b\bar{b})$ and $R_{br}$
        are superposed on the figure.
        Input values are taken to be the same as figure 3(a).
\end{description}

\newpage
%
% FIGURES
%

\begin{figure*}[h]\vspace{-0.3cm}
\epsfxsize = 14 cm
\centerline{\epsfbox{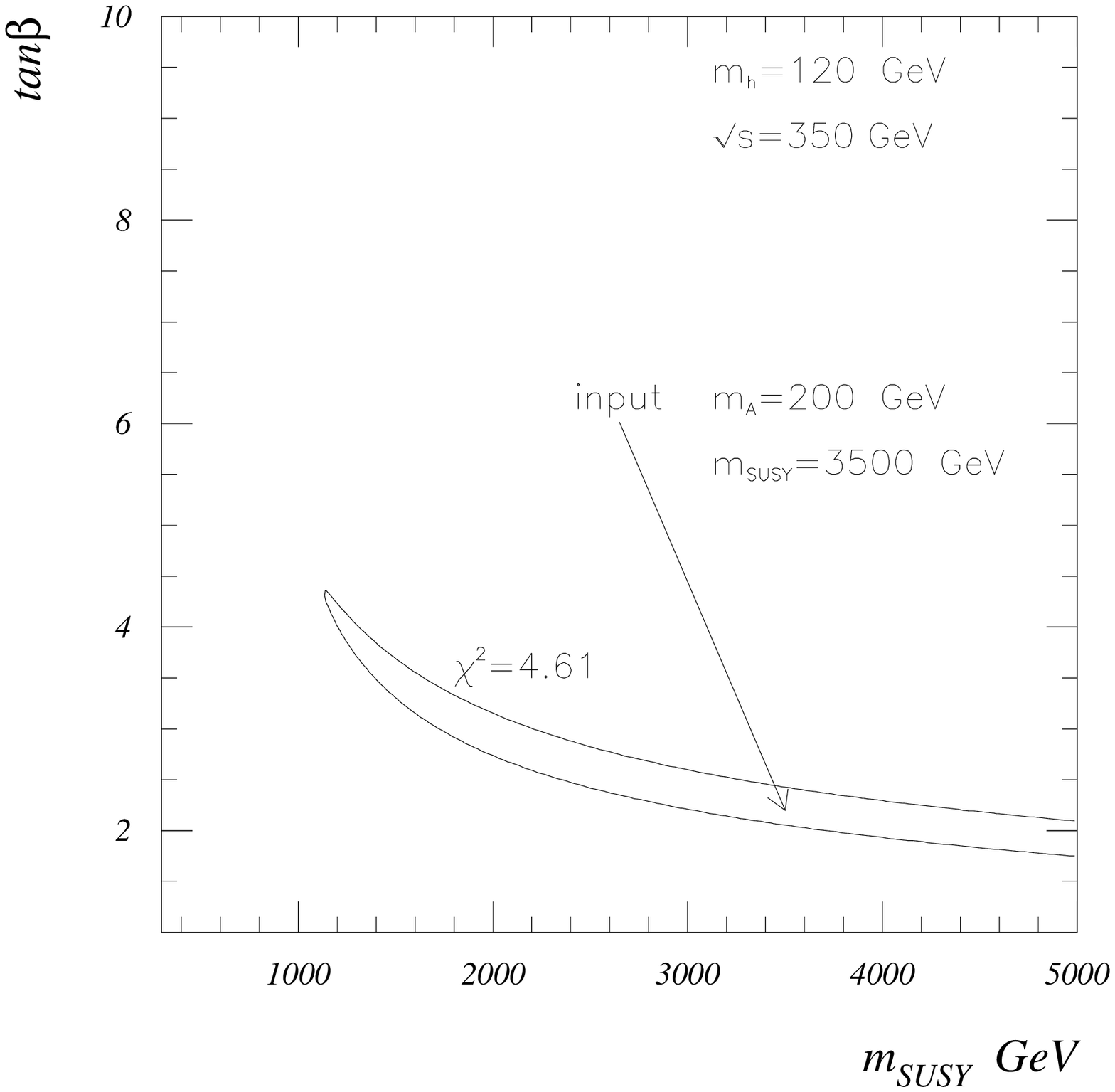}}
\caption[]{(a) }
\end{figure*}
\addtocounter{figure}{-1}
\begin{figure*}\vspace{-0.3cm}
\epsfxsize = 14 cm
\centerline{\epsfbox{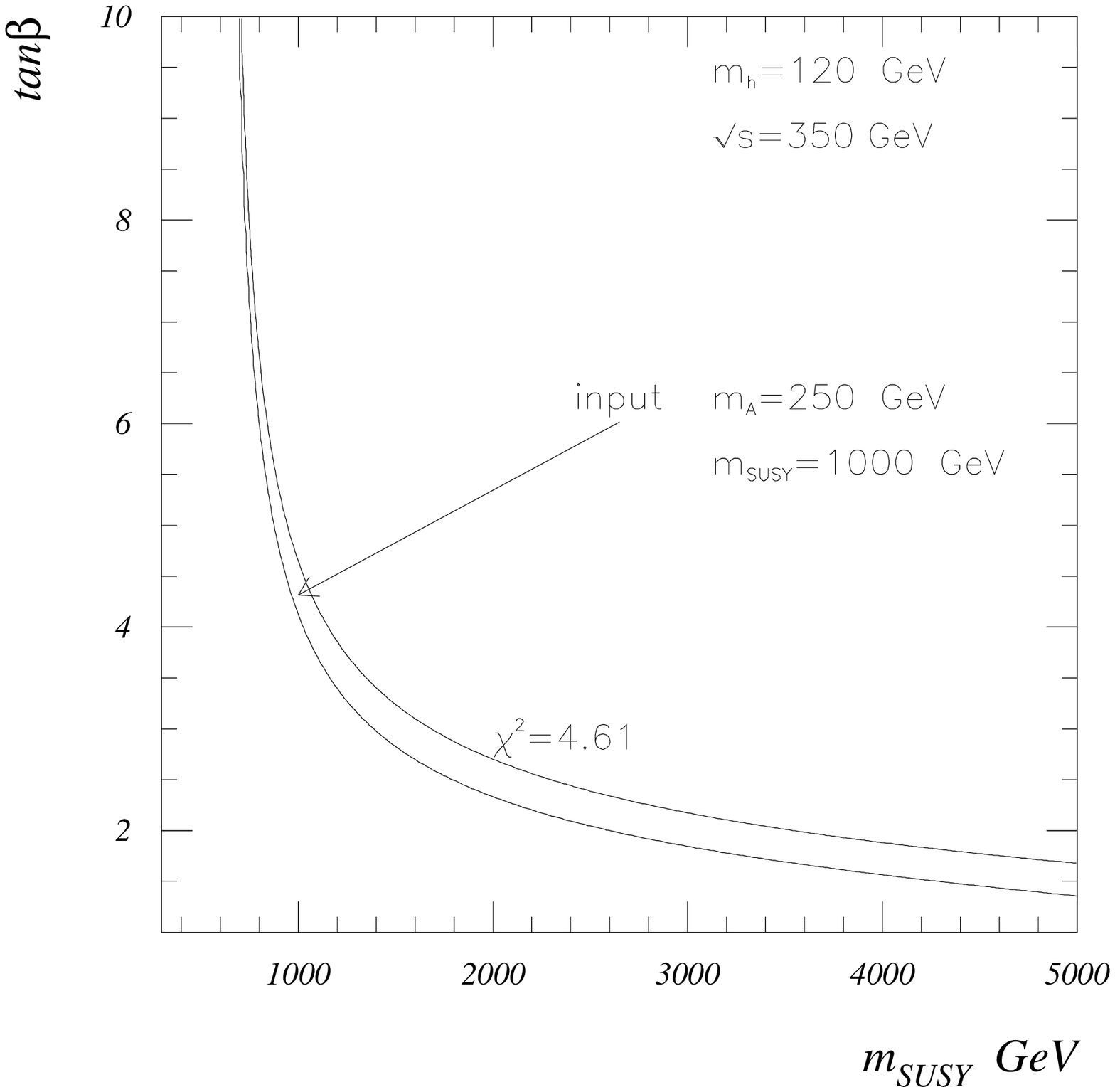}}
\caption[]{(b) }
\label{fign:ch2i}
\end{figure*}

\begin{figure*}\vspace{-0.3cm}
\epsfxsize = 14 cm
\centerline{\epsfbox{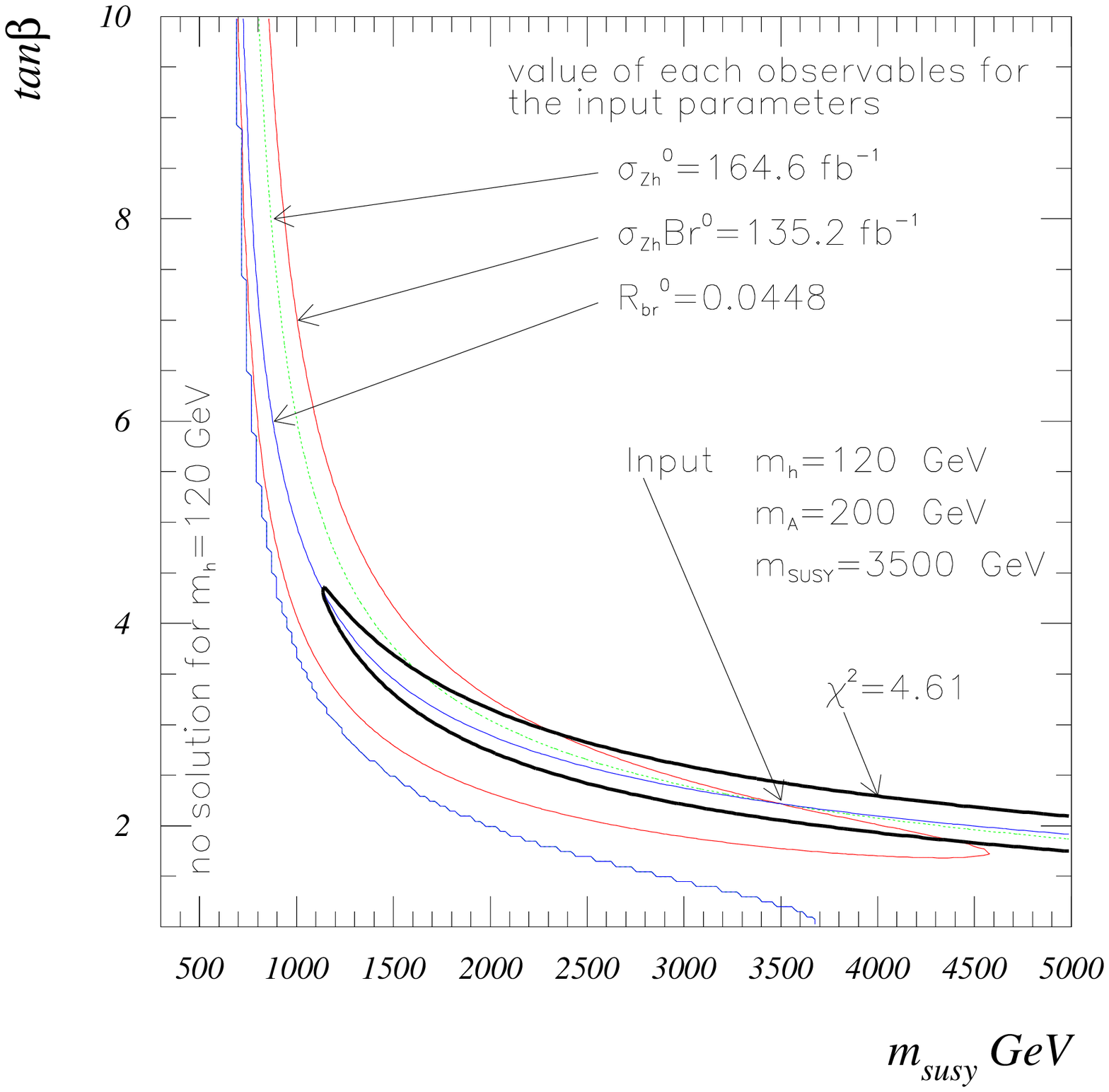}}
\caption[]{ }
\label{fign:ch2spp}
\end{figure*}


\baselineskip 16pt
\begin{thebibliography}{99}
\bibitem{OYY} Y. Okada, M. Yamaguchi and T. Yanagida,
        {\sl Prog. Theor.Phys.}\/ {\bf 85} (1991) 1; 
        {\sl Phys.Lett.}\/ B{\bf 262} (1991) 54;
         J. Ellis, G. Ridolfi and F. Zwirner, 
        {\sl Phys.Lett.}\/ B{\bf 257} (1991) 83;
         H. E. Haber and R. Hempfling,
        {\sl Phys.Rev.Lett.}\/ {\bf 66} (1991) 1815. 
\bibitem{Janot} P. Janot, Workshop,Saariselka,Finland,Sep
        9-14,1991.
        In Munich/Annecy/Hamburg 1991, Proceedings,
        $e^+e^-$ collisions at 500 GeV 107-131
        and Orsay Lin.Accel.Lab L.A.L. 91-61 (91/11,rec.Jan.92) 25. 
        J. Kamoshita,  {\sl INS Workshop, Dec 20-22, 1994.
        "Physics of $e^+e^-$,$e^-\gamma$ and $\gamma\gamma$ 
        collisions at linear accelerators"}, 1994, Proceedings, 
        INS-J-181.
\bibitem{Kamo} J. Kamoshita, Y. Okada, M. Tanaka, 
               {\sl Phys.Lett.}\/ B{\bf 328} (1994) 67.
\bibitem{KOT} J. Kamoshita, Y. Okada, M. Tanaka, 
               {\sl Phys.Lett.}\/ B{\bf 391} (1997) 124.
\bibitem{JLC} JLC-I, KEK Report 92-16, December 1992.
\bibitem{WCHB} See, for example,
          J. F. Gunion, A. Stange and S. Willenbrock
          hep/ph9602238.
\bibitem{NLCHWG} J. F. Gunion et.al. hep/ph9703330,
          To appear in ``Proceedings of the 1996 DPF/DPB
          Summer Study on New Directions for High Energy Physics'',
          UCD-97-5 916 .
\bibitem{HBB} M. D. Hildrech, T. L. Barklow and D. L. Burke
        {\sl Phys.Rev.}\/ {\bf D49} (1994) 3441.
\bibitem{Nakamura} K. Kawagoe, I. Nakamura, 
        {\sl Phys.Rev.}\/ {\bf D54} (1996) 3634 .
\bibitem{Barger} V. Barger, M. S. Berger, A. L. Stange 
                and R. J. N. Phillips, 
                {\sl Phys.Rev.}\/D{\bf 45} (1992) 4128,
        A. Djouadi, J. Kalinowski, P. M. Zerwas,
                {\sl Z.Phys.}\/{\bf C70} (1996) 435-488.
\bibitem{YKWU} B. Pendleton and G. G. Ross 
              {\sl Phys.Lett.}\/B{\bf 98}(1981)291,
        V. Barger, M. S. Berger, P. Ohmann, and R. N. Phillips 
              {\sl Phys.Lett.}\/B{\bf 314}(1993)351, 
        P. Langacker, N. Polonsky, 
              {\sl Phys.Rev.}\/D{\bf 49}(1994)1454-1467,
        S. Kelley, J. L. Lopez and D. V. Nanopulos 
              {\sl Phys.Lett.}\/ B{\bf 274}(1992)387,
        M. Carena, M. Olechowski, S. Pokorski and C. E. M. Wagner 
              {\sl Nucl.Phys.}\/ B{\bf 426}(1994)269.   
\end{thebibliography}
\end{document}